\pdfoutput=1
\documentclass[aps,pra,amssymb,superscriptaddress,nofootinbib,nopreprintnumbers,notitlepage]{revtex4-1}  

\usepackage{graphicx}%,bm}
\usepackage{picture}
\usepackage{placeins}
\usepackage{float}
\DeclareGraphicsExtensions{.pdf,.eps} %.png
\usepackage{amsmath,scalefnt}
\usepackage{amssymb}
\usepackage{verbatim}
\usepackage{amsmath,amsfonts}
\usepackage{amsbsy}
\usepackage{color}
\usepackage{soul}
\usepackage[svgnames]{xcolor} 
\usepackage{soul}

\usepackage[breaklinks=true,colorlinks=true,linkcolor=DarkBlue,urlcolor=DarkBlue,citecolor=DarkBlue]{hyperref}

\usepackage{braket}

\def\ba#1{\left(\begin{array}{#1}}
	\def\ea{\end{array}\right)}
\def\bsm{\left(\begin{smallmatrix}}
	\def\esm{\end{smallmatrix}\right)}

\begin{document}

\title{Comment on ``Interaction of a Bose-Einstein condensate with a gravitational wave''}

\author{Richard Howl}
  \affiliation{School of Mathematical Sciences,
	University of Nottingham,
	University Park,
	Nottingham NG7 2RD,
	United Kingdom}

\author{Dennis R\"{a}tzel}
\affiliation{Institut f\"{u}r Physik, Humboldt-Universit\"{a}t zu Berlin, 12489 Berlin, Germany}

\author{Ivette Fuentes}
\email[Author to whom correspondence should be addressed: ]{ivette.fuentes@nottingham.ac.at}
  \affiliation{School of Mathematical Sciences,
University of Nottingham,
University Park,
Nottingham NG7 2RD,
United Kingdom}

\begin{abstract} 
A gravitational-wave (GW) detector that utilizes the phononic excitations of a Bose-Einstein condensate (BEC) has recently been proposed [NJP 16, 085003 (2014)]. A subsequent and independent study,  [arXiv:1807.07046v1], has suggested an alternative GW detection scheme that also uses phonons of a BEC but which was found to be many orders of magnitude away from  being feasible. Here we make clear that the two proposed schemes are very different and that the conclusions of  [arXiv:1807.07046v1] do not apply to the original proposal [NJP 16, 085003 (2014)].  
\end{abstract}
 
\maketitle

In \cite{Ralph}, the author studies a gravitational-wave (GW) detection scheme that is based on phonons of Bose-Einstein Condensates (BECs) but which is found to be impractical.  Since a previous proposal \cite{GWDetectorFirst} for GW detection also used phonons of a BEC, it could be interpreted that  the conclusions of \cite{Ralph} also apply to \cite{GWDetectorFirst}. However, there are many differences between these proposals, and here we present the two most important ones:  i) despite the reference to squeezing in the abstract of \cite{Ralph}, the initial phononic field is not squeezed, in contrast to the original proposal \cite{GWDetectorFirst}; and ii) no trapping potential is considered in \cite{Ralph}, which is vital to any experimental realization of a BEC, and was found to be critical to detecting GWs in \cite{GWDetectorFirst}. 

Addressing point i) first, an integral part of the proposal \cite{GWDetectorFirst} is the use of quantum metrology.  
Here one considers an initial state, also called a probe state (for example,
the vacuum or a squeezed state),  a transformation that encodes the parameter of interest (in
this case the GW strain $\epsilon$) and, finally, a measurement. In order to estimate the parameter with high precision, it is necessary to distinguish two states  $\ket{\psi(\epsilon)}$ and  $\ket{\psi(\epsilon + d \epsilon)}$ that differ by an infinitesimal change  $d \epsilon$ of the parameter  $\epsilon$. The operational measure that quantifies the distinguishability of these two states is the Fisher information, and optimizing over all the possible quantum measurements provides the quantum Fisher information (QFI). The QFI depends strongly on the nature of the transformation that encodes the parameter and on the initial quantum state. In particular, it can be shown that, in a single-run of an experiment, the error in the estimation of the parameter of interest can scale as $1 / N$, where $N$ is the number of initial probes (in this case phonons) for particular quantum states, rather than $1/\sqrt{N}$ as for classical states.  The optimum $1/N$ scaling, called Heisenberg-scaling, can be achieved with squeezed states \cite{SqueezedHeisenberg1,SqueezedHeisenberg2}, and these are the initial states prepared in \cite{GWDetectorFirst} but not in \cite{Ralph}. For an initial squeezed phonon state with $N \gg 1$,  the error in the estimation of the GW strain is \cite{GWDetectorFirst}:
\begin{align} \label{eq:squeezed}
\Delta \epsilon \propto \frac{1}{ N \sqrt{\omega_m \omega_n \tau t N_d}},
\end{align}
where $t$ is the time of each experimental run; $\tau$ is the integration time; $N_d$ is the number of detectors; and  $\omega_m$ and $\omega_n$ are the angular frequencies of two phonon modes, which can be related to the GW angular frequency, $\Omega$, through either $\Omega = \omega_m + \omega_n$ or $\Omega = \omega_n - \omega_m$. The former relation holds when a parametric resonance process occurs, whereas the latter is a resonant frequency-converting (mode-mixing) process \cite{GWDetectorFirst}. Note that, in the latter case, the phonon frequencies can be greater than the GW frequency. 

 With large phonon numbers, a squeezed state can lead to orders of magnitude improvement over using a coherent or vacuum state where $N$ in \eqref{eq:squeezed} would be replaced by $\sqrt{N}$ in the former case, and approximately $1$ in the latter case.  By mixing the condensate and phononic modes as studied in \cite{PhononSpectroscopy,PhononEvap}, it is possible to further improve this sensitivity for a given experimental setup since the factor $N$ is essentially now replaced with $\sqrt{N_0 N}$ where $N_0$ is the number of condensate atoms \cite{ResonanceInter}, which must be greater than the number of phonons for the description of the Bose gas used in \cite{GWDetectorFirst,Ralph}.  Note that, from \eqref{eq:squeezed}, it is not one single number that will determine how good  the estimation is; rather, it is a combination of several factors.  As often occurs in quantum metrology, the error in the estimation is also independent of the parameter of interest $\epsilon$, which is the strain of the GW in our case. Unfortunately, the error in the estimation of the GW strain is not calculated in \cite{Ralph} for their scheme, instead general arguments are provided for why it is expected that it would not currently be able to detect GWs.

We now address point ii). In all BEC experiments, the Bose gas is trapped, and this is clearly   fundamental to the experimental system. When assessing how a GW interacts with an experimentally viable BEC, we must consider how the trap is affected by a GW. However, in \cite{Ralph} no  trap potential is assumed. This fact is also emphasized by the author in the conclusions section of \cite{Ralph}. There they also  shortly discuss the implications of a trapping potential and state ``However, all the arguments above would still apply''. The validity of this statement would need to be shown in detail. In contrast to \cite{Ralph}, in \cite{GWDetectorFirst}, a uniform box trap potential is considered, which has been used in experiments \cite{BoxTrap}.  The device creating the tapping potential can be considered to be rigid to the GW in comparison to the BEC.   To describe this situation, it is assumed that the proper length remains constant, which is a frame-independent concept (see \cite{OpticalRes} for an introduction and an application to an optical resonator). However, we must be very careful to be consistent in whichever frame we choose to describe the process. In \cite{GWDetectorFirst}, the transverse-traceless (TT) frame is used for the GW. In this frame the coordinate extension of the trap will  be changing, leading to moving boundary conditions and the resonance effects that are used for the detection scheme of \cite{GWDetectorFirst}.  Alternatively, one can  use a frame closer to the lab frame \cite{ProperDetectorFrame,Thorne1983} (often referred to as the  proper detector frame) where the  coordinate extension of the trap stays fixed.  In this case, the GW effect enters (to first order) as an additional potential term in the Gross-Pitaevskii equation. This can be derived by considering a Lagrangian  for a covariant BEC (see e.g.\ \cite{RelBECs}) in a GW spacetime  and then taking the non-relativistic limit of a BEC \cite{GWSeismic} or by simply considering that the GW essentially acts as a Newtonian force in this reference frame \cite{GWBook}. In this case, it is the additional  potential that leads to  the  resonance effects that are used for the proposal \cite{GWDetectorFirst} (see also \cite{MassPaper} for a similar process deriving from a moving mass close to a BEC, where a quantum and classical treatment of phonons is employed).  In addition to these parametric processes, there is also a direct amplification channel, but this has a much smaller effect on the phonons \cite{MassPaper}. 

In \cite{Ralph}, the TT frame is used for describing the GW but no trap potential and, therefore, no boundary conditions are considered, leading to a too simplistic description of how a GW interacts with experimental BECs.  In particular, it is  explained that a GW acting on a homogeneous BEC at rest does not create phonons and instead only changes the propagation of phonons, with squeezing  being a higher order effect.  While this seems to be correct for homogeneous BECs without boundaries, a box trap potential is considered in \cite{GWDetectorFirst} and the  arguments presented above explain why a  GW acting on a box-trapped BEC  at rest does indeed create phonons, and that this is not a higher order effect.\footnote{Note that in the Thomas-Fermi limit of a boxed-trapped BEC, the phonons will effectively see a homogeneous condensate, similar to the case considered in \cite{Ralph}, but, importantly, with  boundary conditions  \cite{PitaevskiiBook,MassPaper}.}  An inhomogeneous case is also discussed in \cite{Ralph} where GWs could directly create phonons. However, the BEC is still not in a trapping potential, and this is very different to the case considered in the original proposal \cite{GWDetectorFirst}. Furthermore,  while a linear dependence of the fidelity on the GW strain is discussed in \cite{Ralph}, which is also a feature in \cite{GWDetectorFirst},  no parametric amplification mechanism is considered, whereas these processes form the detection mechanism of \cite{GWDetectorFirst}. 

 Although point i) and ii) above are the two main differences between the proposal of \cite{Ralph} and \cite{GWDetectorFirst}, there are several other distinguishing points. For example, the possibility of resonances is only shortly mentioned in \cite{Ralph} and is not taken into account seriously, while resonances are central in \cite{GWDetectorFirst}. Also, the time scale of a single experiment $\tau$ and the total duration of the measurement $t$, both appearing in equation (\ref{eq:squeezed}), are not considered in detail in \cite{Ralph}. Instead a comparison is drawn with LIGO-like interferometric GW detectors, which usually measure transient phenomena that only last for milliseconds, while the mechanism of \cite{GWDetectorFirst} has only been proposed for persistent sources that may allow for much  larger integration times. 

In conclusion, we have discussed how the principle components of the original proposed detection scheme \cite{GWDetectorFirst} -  quantum metrology with an initial squeezed state of phonons, and using the fact that experimental BECs are contained within trapping potentials - were not considered in the detection scheme proposed in \cite{Ralph}, resulting in much poorer  sensitivity for that scheme compared to \cite{GWDetectorFirst}. Using an initial squeezed state results in the optimum scaling, Heisenberg scaling, in sensitivity, and the boundary conditions due to the trapping  potential result in the direct  resonant squeezing or mode-mixing   of phonons, which is essential to the detection scheme of the original proposal. 

\textbf{Note Added}: After this work was completed, we became aware of a new version of \cite{Ralph}, [arXiv:1807.07046v2]. This comment is for the original version [arXiv:1807.07046v1].

\section*{Acknowledgements}

We thank David E. Bruschi  for useful discussions and comments. RH and IF  would like to acknowledge that this work was made possible through the support of the Penrose Institute and the grant `Leaps in cosmology: gravitational wave detection with quantum systems' (No. 58745) from the John Templeton Foundation. The opinions expressed in this publication are those of the authors and do not necessarily reflect the views of the John Templeton Foundation. DR thanks the Humboldt Foundation for funding his research with their Feodor-Lynen Fellowship. 

\newpage

%\bibliography{references}
\bibliographystyle{apsrev4-1}

\end{document}